\documentstyle[prl,preprint,aps,epsf,tighten]{revtex}
\draft
\preprint{HKBU-CNS-9829}
\begin{document}
\title{Numerical accuracy of Bogomolny's semiclassical quantization scheme
in quantum billiards}
\author{Bambi Hu$^{1,2}$, Baowen~Li$^{1}$, and Daniel~C~Rouben$^{1}$}
\address{$^{1}$ Department of Physics and the Centre for Nonlinear Studies,
Hong Kong Baptist University, Hong Kong, China}
\address{$^{2}$ Department of Physics, University of Houston,
Houston, TX 77204-506, USA}
\date{\today}
\maketitle

\begin{abstract}

We use the semiclassical quantization scheme of Bogomolny to calculate
eigenvalues of the Lima\c con quantum billiard corresponding to a
conformal map of the circle billiard. We use the entire billiard boundary
as the chosen surface of section (SOS) and use a finite approximation for
the transfer operator in coordinate space. Computation of the eigenvalues
of this matrix combined with a quantization condition, determines a set of
semiclassical eigenvalues which are compared with those obtained by
solving the Schr\"odinger equation. The classical dynamics of this
billiard system undergoes a smooth transition from integrable (circle) to
completely chaotic motion, thus providing a test of Bogomolny's
semiclassical method in coordinate space in terms of the morphology of the
wavefunction. We analyze the results for billiards which exhibit both soft
and hard chaos. 

\end{abstract}


\section{Introduction}

This paper presents a numerical investigation of Bogomolny's 
semiclassical scheme for solving the quantum problem of a 
non-integrable billiard system. We deal with a single particle in a 
zero field environment making specular collisions with a closed 
and singly-connected boundary. The analogous quantum problem refers 
to the solution of the Helmholtz equation in a closed
region {\it B},
\begin{equation}
\Delta_{\vec r}\psi(\vec r)+k^2\psi(\vec r)=0
\label{Schr}
\end{equation}
with Dirichlet boundary condition, $\psi(\vec r)|_{\partial B}=0$ and
$\hbar=1$, $2m=1$. Such a system is called a quantum billiard.

During the past twenty years a tremendous effort has been devoted to 
the study of quantum systems whose classical counterpart is chaotic. 
A detailed analysis of the results comprises the statistical 
properties of the energy spectrum and the geometric structure or 
morphology of the eigenfunctions and their statistics. Some recent 
reviews with many references to the progress made can be found in 
Gutzwiller's book \cite{GutzB}, Giannoni {\it et al.} \cite{BohigasLesH91} 
and Casati and Chirikov \cite{Casati93}. At the heart of these 
studies lies the issue of quantum chaos or the influence of classical
chaos on the solutions of Eq.~(\ref{Schr}). On the one hand, past research
has considered the statistics of the energy-level spectrum and has
demonstrated the existence of universal classes for particular classical 
regimes. Integrable dynamics lead to uncorrelated energy levels (Poisson 
spectrum) and completely chaotic systems lead to Wigner-Dyson statistics of 
one of the standard ensembles of random matrices. On the other hand 
the notion of scars has played an important role in the study of eigenfunction
structure 
\cite{Heller84,Bogomolny88,Berry89,Fishman93,Li97,Li98,Heller98}.

As for the methods to solve the quantum problem, many have been proposed and
the most successful will be quickly mentioned. Most textbooks deal
with integrable systems since an analytic solution for the energy exists
and can be easily written down. Unfortunately, they do not even mention 
generic non-integrable chaotic classical systems for which the quantum 
problem can only be solved numerically. Of all the methods, two have stood 
out to be particularly successful; they are the plane wave decomposition 
method (PWDM) invented by Heller\cite{Heller84,HPWDM} and improved by Li
{\it et al.}\cite{Li97,Li98,Li94}, and the boundary integral method
(BIM) \cite{Berry84,Boasman,LRH98}.

Other methods also deserve mention. Firstly we consider the conformal map
diagonalization technique, which was first used by Robnik to calculate 
the eigenvalues of the Lima\c con billiard \cite{Robnik83B}. This method 
makes use of a conformal map which transforms the boundary to an integrable 
geometry but adds additional terms to the Hamiltonian. This method was 
later invoked by Berry and Robnik\cite{Berry86}, Prosen and Robnik 
\cite{PR93} and Bohigas {\it et al.}\cite{Bohigas93}. In addition, the 
scattering quantization method introduced by Smilansky and co-workers 
\cite{uzy1} provides an alternative approach for solving the eigenvalue 
problem. Prosen \cite{Prosen96} has extended this scattering quantization 
method to an exact quantization on the surface of section. In his 
method, the exact unitary quantum Poincar\'e mapping is constructed 
quite generally from the scattering operators of the related scattering 
problem whose semi-classical approximation gives exactly Bogomolny's 
transfer operator which shall be studied exclusively in this paper. 
As for the method of calculating high-lying eigenstates, we would like 
to mention the method introduced by Vergini and Saraceno\cite{VS95}. 
This method overcomes the disadvantage of missing levels via the PWDM 
and in the BIM, and can directly give all eigenvalues in a narrow 
energy range by solving a generalized eigenvalue problem. However, 
it is still not clear whether this method applies to a billiard 
having nonconvex boundary.

The above methods attempt to find solutions of the Schr\"odinger equation
by an exact albeit numerical method and as a result a small error must be
expected. Another approach, not exact, refers to quantization in the
semiclassical limit, $\hbar\to 0$, or quantization of high lying
eigenvalues. The study of quantum chaos and the extraction of eigenvalues
via semiclassical methods began with the quantization of integrable systems 
or EBK theory. In the past forty years, though, much effort has been spent to 
extend the semiclassical approach to quantum systems whose classical 
counterpart is not integrable. Naturally, this study is intimately 
tied to the study of quantum chaos.

The first semiclassical approximations for classically chaotic systems are
all based on the Gutzwiller trace formula (GTF) \cite{GutzB}. In this theory,
the density of state ($\rho(E) =\sum_i \delta(E-E_i)$) is expressed as a 
sum of two terms, the first being the usual Weyl formula or Thomas-Fermi 
density of states and the second being a long range surface correction 
term. In 2-D one writes:
\begin{equation}
\rho(E)\approx \bar{\rho}(E) + \Im m
\frac{i}{\pi\hbar}\sum_p\sum_{r=1}^{\infty}
\frac{T_p(E)}{\sqrt{|det(M^r_p(E)-I)|}}
\exp {\left[(\frac{i}{\hbar}S_p(E) - i\frac{\pi}{2}\mu_p)r\right] }
\label{GTF}
\end{equation}
where the second term on the right hand side contains information 
pertaining to primitive periodic orbits and their repetitions, 
labeled by $p$. This includes the action $S_p$, the trace of the
monodromy matrix $M^r_p$, a phase factor, $\mu_p$ and the geometrical 
period $T_p(E)=\frac{\partial S_p}{\partial E}$. While formally
applicable for systems whose classical counterparts are either chaotic,
integrable or a mixture of the two, the GTF (in its form written above)
has been shown to fail, or at best proved difficult to implement for two 
independent reasons. One problem deals with mixed systems where the main 
contribution from bifurcating periodic orbits is divergent, 
$TrM_p\to\pm 2$, and has motivated a reevaluation of each orbit's 
contribution close to the bifurcation. The other deals with the 
divergence of the sum due to an exponential proliferation of 
classical orbits (e.g. an exponential increase in the number of 
long-range periodic orbits with energy). In recent years, there have 
been many works devoted to overcome these two shortcomings of the 
GTF. On the one hand, the cycle expansion method \cite{cycleexp} and an 
energy-smoothed version of the GTF \cite{Aurich} have been invented to 
provide a numerically efficient and convergent method to evaluate 
periodic-orbit expressions. On the other hand, many works have 
been down to the addition of higher terms of $\hbar$\cite{GTF2}.

More recently another approach to the semiclassical quantization problem was
presented by Bogomolny \cite{Bogomolny92}. Bogomolny's method makes use of a 
finite representation of a quantum Poincar\'e map in the semiclassical limit
and leads to the calculation of a matrix whose eigenvalues allow one to 
determine an approximation to the true energy eigenvalues. It is founded 
on the BIM for determining wavefunctions inside a billiard system and its 
theoretical motivation came from two facts
\begin{itemize}
\item one, for generic billiard problems in any kind of external field, 
there does not exist an explicit closed form solution for the Green 
function; 
\item  two, for general boundary condition involving $\psi$ and its 
derivatives
one cannot reduce the solution of Eq.~(\ref{Schr}) to an integral equation
which can be easily solved numerically as in typical problems where the BIM
is useful.
\end{itemize}
Bogomolny worked around this by employing the standard semiclassical
formula for the Green function in the energy representation and working 
in a boundary integral-like setting. He introduces by this way, a 
semiclassical transfer operator $T(E)$ and a quantization condition 
for an eigenvalue ${\bf det}(I-T(E))=0$.

Some past applications of Bogomolny's method, include the following 
examples. It has been successfully applied to the rectangular billiard 
\cite{Lauritzen}, and to billiards with circular symmetry 
\cite{Tong,Goodings97,Goodings98}. All of these systems are integrable. 
For systems exhibiting hard chaos, we can mention applications to the 
geodesic flow on surfaces of constant negative curvature 
\cite{Bogomolny92B} and to the wedge billiard \cite{GoodingsN}. An
interesting study \cite{Haggerty} has also been carried out by applying
Bogomolny's transfer operator to a smooth nonscalable potential, the 
Nelson potential, at two fixed energies which correspond one to motion 
that is mostly integrable and the other mostly chaotic. However, to date 
an exhaustive test of the method for various classical regimes has not 
been done or at least not to very high energies. Goodings {\it et al.} 
\cite{GoodingsN} explore the first 30 eigenvalues for angles of the 
wedge billiard corresponding to soft chaos.

In this paper we report the results of Bogomolny's semiclassical 
quantization scheme in a closed billiard whose boundary is derived from 
the quadratic conformal map of the unit circle \cite{Robnik83A}, namely 
the Lima\c con billiard. The mapping is controlled by a single parameter 
$\lambda$, with $\lambda=0$ corresponding to the circle billiard:
\begin{eqnarray}
x&=&\cos(\theta)+\lambda\cos(2\theta)\nonumber\\
y&=&\sin(\theta)+\lambda\sin(2\theta).
\label{boundary}
\end{eqnarray}
For all $\lambda,\; 0\le\lambda <1/4$ the boundary is analytic but 
non-convex for $\lambda>1/4$. The classical dynamics of this system 
has been extensively investigated\cite{Robnik83A}, and shown to 
undergo a smooth transition from integrable motion, $\lambda=0$ to 
a soft chaos, KAM regime, $0<\lambda\le 1/4$. At the convex-concave 
transition point, $\lambda=1/4$, the motion is very nearly ergodic 
\cite{Robnik83A}. While Hayli {\it et al.} \cite{Hayli87} have shown that 
some very small stable islands still exist in the phase space, it 
can be supposed that above a $\lambda\approx 0.28$ these stable islands 
also disappear and the dynamics be that of hard chaos with mixing and 
positive K-entropy. A particular case is $\lambda=1/2$, for which the 
billiard boundary has one non-analytic point and it has been rigorously 
shown \cite{Markarian}, that the motion exhibits hard 
chaos. For $\lambda=0.15$ the chaotic regions cover
$64.6\%$  of the phase space and for $\lambda=0.2$ they cover more
than $90\%$ \cite{PR93}.

To understand, at least, from a qualitative point of view, how Bogomolny's
semiclassical method might work in the Lima\c con billiard we may consider
three independent factors: (1) the success of the BIM without any
semiclassical approximation; (2) the effects that a transition through a
mixed regime will have on the morphology of each eigenstate; (3) the 
propensity of Bogomolny's method for describing the energy of any particular
kind of eigenstate based on the classification scheme, regular, mixed, and
chaotic. The latter classification was first proposed by Percival
\cite{percival} and used as the essential ingredients for the energy 
level statistics in mixed systems, namely the Berry-Robnik surmise
\cite{BerryRobnik}. Moreover, a detailed classification of states as 
being either regular or chaotic in the deep semiclassical limit has
been performed with great success for the system studied here
\cite{Li95A,Li95B} and also for much higher energy levels in another
system\cite{Prosen96}. However it is clear that this may not be possible at
lower energies, the first 200 states, where the effective $\hbar$ occupies
a phase space area $\propto 1/\sqrt E$ and is too large for sufficient
resolution of projected states on a Poincare surface of section, 
(see Sec. 3). In this paper we will consider values of $\lambda$ 
corresponding to soft chaos, $\lambda=0.1, 0.15$ and to harder chaos 
(or nearly hard chaos), $\lambda=0.2, 0.23, 0.245, 0.25$. 
At $\lambda=0.15$ we also study the method in the deep semiclassical 
limit where a classification of states is possible. 

As was stated above, the function, $det(I-T(E))$, equals zero as $E$ equals an 
energy eigenvalue. Most importantly, though, a formal relation between this 
determinant and the GTF has been demonstrated \cite{Bogomolny92}
$$\ln\left({\bf det}(I-T^{osc}(E))\right)=
\sum_p \frac{T_p}{\hbar\sqrt{|det(M_p-I)|}}
\exp{\left(\frac{i}{\hbar}S_p-i\frac{\pi}{2}\mu_p\right)}$$
so that formally the zeros of the Bogomolny Transfer Operator $I-T(E)$
must be close to the positions of the poles of the GTF (the zeros of the
former are separated from the poles of the latter by one stationary phase
integral \cite{Bogomolny92}). While strictly formal, this relation does
beg a comparison between the two methods in the purely hard chaos regime
and some very good results have been obtained using many thousands of 
unstable periodic orbits in strongly chaotic billiard systems \cite{Aurich}. 
But again, these results were obtained using methods that are most 
appropriate to systems exhibiting hard chaos and as such are not readily
applicable to the system at hand.

\section{The transfer operator in coordinate space}

We will deal with a closed billiard system with the Dirichlet boundary 
condition, in which the transfer operator of Bogomolny acts as a quantum 
Poincar\'e map \cite{Prosen96}. We have mentioned that Bogomolny's method
is founded on the BIM. This interpretation however is not necessary. We
may also derive Bogomolny's quantization condition by starting with the
general theory of quantum Poincar\'e maps (QPM). In this theory, one must
initially define a certain surface of section (SOS), {\it P}, a set of 
coordinates $s$ on ${\it P}$ and a domain, ${\it L}$, for the QPM: in 
other words a set of ${\it L_2}$ functions in ${\it P}$.

Prosen \cite{Prosen96} considers a certain non-unitary, compact QPM 
constructed from the product of two scattering transfer operators and 
shows that its semiclassical limit reduces to the transfer operator 
of Bogomolny. In general a QPM acts on $\psi_1\in {\it L}$ to give a 
$\psi_2\in {\it L}$. 
In the coordinates representation on ${\it P}$ this is given by,
\begin{equation}
\psi_2(s')=\int_{SOS}ds T(s,s',E)\psi_1(s)
\label{condition}
\end{equation}
A viable quantization condition (i.e. one that corresponds to the Schr\"odinger
equation of the full system) can be obtained by requiring that for
some energy, $E^*$, there exists some $\psi$ in {\it L} that is left unchanged
after one Poincar\'e mapping \cite{Prosen96}. This means that the QPM has at
least one unit eigenvalue at $E^*$.

For generic problems one can construct a special scattering Hamiltonian in
the SOS and from this an exact non-unitary compact QPM as in
\cite{uzy1,Prosen96}. On the other hand, to study the problem semiclassically
one replaces the same QPM with its semiclassical limit which is
Bogomolny's transfer operator, $T^{osc}$ (from now on we denote the
transfer operator of Bogomolny by $T^{osc}$ to distinguish it from any
other QPM). Using Eq.~(\ref{condition}), this leads to the familiar 
semiclassical quantization condition \cite{Bogomolny92}, 
\begin{equation}
{\bf det}(I-T^{osc}(E)) = 0
\label{BOGQC}
\end{equation}
where $T^{osc}$ is constructed from classical trajectories passing from the 
SOS back to itself after one application of the QPM. The paper \cite{Prosen96} 
also indicates how to treat higher order semiclassical errors for the 
energy (please also see \cite{Boasman,GTF2} and references therein).

In solving the equation (\ref{BOGQC}) we must first express the
transfer operator in a finite dimensional basis and the roots
are then computed numerically. A particularly convenient choice 
for the SOS is the billiard boundary with a basis formed from a 
discretization of the boundary in either momentum or coordinate 
space. We have chosen the coordinate space representation and a 
discretization of the boundary into $N$ cells of length
$\Delta_m (m=1,...,N)$. Let $s$ be the coordinate that measures distance
along the boundary, and let $s=s_n$, $s'=s_m$ be two points in two
different cells. The semiclassical description for calculating the 
matrix element, $T^{osc}_{n\;m}$, is to sum over all possible 
classical trajectories which cross the SOS at $s_n$ and $s_m$
after one application of the Poincar\'e map\footnote{We note
that the SOS is really a small distance $\epsilon$ from the boundary.},

\begin{equation}
T^{osc}(E)=
\left\{
\begin{array}
{ll}\frac{-1}{(2i\pi\hbar)^{1/2}}|\frac{\partial^2 S(s_n,s_m;E)}
{\partial s\partial
s'}|^{1/2}\exp{\left[\frac{i}{\hbar}S(s_n,s_m;E)-i\frac{\pi}{2}\nu\right]}
\sqrt{\delta_n\delta_{m}}& n\ne m
\\1.0\; &n=m.
\end{array}
\right.
\label{element}
\end{equation}

By using the entire boundary we have the numerical simplification that each
cell is connected by one unique trajectory whose action $S(s_n,s_m;E)$
at energy $E$ is the length of the chord passing between these points
multiplied by the factor $\sqrt{E}$. For the Dirichlet boundary
condition and a convex geometry we put the phase index $\nu$ equal to 2 for
all matrix elements. For the Dirichlet boundary condition and a non-convex
geometry one can include ghost trajectories \cite{Bogomolny92} that go outside 
the billiard to connect cells. Finally the prefactor in Eq.~(\ref{element}) 
contains the mixed second derivative of the action, $|\partial^2 S/\partial 
s\partial s'|^{1/2}$. The latter can be conveniently related to the linear 
Poincar\'e map for going from the boundary back to itself.

Bogomolny also writes down a prescription for the dimension of the transfer
matrix which says that in passing from an operator to a finite representation 
one should give the $T$ matrix a dimension no smaller than the number given by:
\begin{equation}
dim(T)\ge {\it A}(E)/(2\pi\hbar)={\cal L}\sqrt{E}/\pi,
\label{minimum}
\end{equation}
where ${\it A}$ is the classically allowed area, ${\cal L}$ is the
billiard perimeter and $\hbar=1$. One could study the curve 
$f(E) = {\bf det}(I-T^{osc}(E))$ but we prefer to examine individual 
eigenvalues of the $T$ matrix. In practice we do not exactly satisfy 
the quantization condition, only in the limit of an infinite transfer 
matrix can the unitarity condition of Bogomolny's transfer operator be
recovered and an eigenvalue of the matrix be exactly one. In this case
we are obtaining an exact solution to Eq.~(\ref{BOGQC}).

Finally we note that the dimension of the $T^{osc}$ matrix is also equal 
to the number of cells on the SOS. Assuming that all cells have the same 
size we then define a parameter $b$, \cite{Boasman,LRH98}, which 
represents the number of cells that make up one de Broglie wavelength, that is
$b=2\pi dim(T)/(\sqrt{E}{\cal L})$. Or by using Eq.~(\ref{minimum}) this
corresponds to $b$ being no smaller than two $2$ (at least $2$ cells per
wavelength). Of course we should and must use a higher dimension and our
calculations were performed with matrices exceeding the number in the Weyl
formula for a one freedom by a factor of 10-20. In the very deep semiclassical
limit, we were not able to go beyond $b\approx 5$ due to the limitation 
of computation facilities.

\section{Numerical calculation of eigenenergies}

The boundary of the Lima\c con billiard is given by Eq.~(\ref{boundary}).
For $\lambda=0$ the eigenenergies are the zeros of the Bessel functions.
Let us denote the $nth$ zero of the $mth$ order Bessel function by 
$\chi^i_j$. The quantum energies are given by the square of this for 
$\hbar=1$ and $m=0.5$. For $\lambda>0$ each zero $\chi^i_j (j>0)$ splits 
into an eigenstate of odd/even parity with respect to reflection on the 
x-axis. The $j=0$ order zeros, on the other hand, are associated with 
even states, so they do not satisfy the Dirichlet boundary condition 
on the x-axis.

In choosing an appropriate coordinate space basis, we have several choices:
\begin{itemize}
\item
By choosing the entire billiard boundary as a basis and putting $\nu=2$
in Eq.~(\ref{element}) for all matrix elements we quantize implicitly all 
states which satisfy the Dirichlet condition for these values of $(x,y)$. Then 
since both odd and even states satisfy this condition, without any reference to 
the latter being zero also on the x-axis, we will obtain a spectrum for both 
symmetry classes.
\item
By choosing the upper boundary and x-axis as a basis and choosing $\nu=2$
for all matrix elements in Eq.~(\ref{element}) we will obtain a quantization 
for only odd states.
\item
By choosing a basis consisting of only half the boundary and constructing
an appropriate symmetry desired Green function by placing a hard (resp. soft) 
wall on the x-axis we obtain a quantization condition for odd (resp. even) 
eigenvalues. In this case, however, we must include one or two trajectories 
going between each cell.
\end{itemize}

When the boundary is non-convex, $\lambda>0.25$, one must be careful in how to
connect cells. In (i), certain pairs of cells cannot be 
connected. We then have 
the choice of putting that matrix element to zero or including the so-called
ghost orbit \cite{Bogomolny92}. In the latter case, however, not all matrix 
elements will have the same factor $\nu$. Noting that the matrix must be 
symmetric leads to the rule that the number $\nu$ is either zero or two 
depending on whether the trajectory crosses the SOS an even or odd number of 
times. For the purely convex case, $\lambda\le 0.25$, all trajectoris cross 
the SOS only once after one Poincar\'e map and $\nu=2$. In (iii) 
we must 
decide which cells have one or two orbits connecting them; obviously if only 
a ghost orbit can connect two cells, there is only one trajectory, otherwise 
there are two trajectories. In order to compute only odd states, we can put 
a hard wall along the x-axis and consider a basis consisting of the x-axis 
and the upper boundary as in (ii). In (ii) one must consider only one 
trajectory connecting two cells. However, the odd state eigenvalues computed 
this way were found to be a factor of 10 worse than using choice (i)
\footnote{This result was also found to be true in the stadium billiard 
considered later. Here we have studied the odd-odd parity symmetry class by 
using both the entire stadium and the quarter stadium. We find, in accordance 
with the Lima\c con billiard, that the results are closer to the reference
eigenvalues when using the entire boundary. The factor is nevertheless greater. 
Using the quarter stadium we find a precision of about 15 percent of the mean 
level spacing whereas the full stadium gives a precision of roughly 1 
percent of the mean level spacing for eigenstates of odd-odd parity. These 
results accidentally coincide with those found by using the BIM 
\cite{LRH98}.}.

For $\lambda<0.25$ we find absolutely no contradiction with using (i) and 
all results presented in figures are based on this choice. Naturally we 
choose $\Delta_i=\Delta={\cal L}/dim(T)$ for all cells and from
Eq.~(\ref{boundary}) we calculate the billiard's area and perimeter. The
latter, ${\cal L}$, is given by $4(1+2\lambda)E(8\lambda/(1+2\lambda))$ 
where $E(k)$ is the complete elliptic integral of the second kind. The 
half boundary perimeter plus the x-axis then has length, 
${\cal L}_{1/2}=\frac{\cal L}{2}+2$. As for the billiard area ${\cal A}$, 
it is $\pi(1+2\lambda^2)$. We then construct a formula for the number of 
odd and even states below some energy $E$, from the usual semiclassical 
formulae \cite{PR93},
\begin{eqnarray}
N_{odd}(E)&=&\frac{\cal A}{8\pi}E-\frac{{\cal L}_{1/2}}{4\pi}\sqrt{E}+
\frac{5}{24}
\nonumber\\
N_{even}(E)&=&\frac{\cal A}{8\pi}E-\frac{{\cal
L}-{\cal L}_{1/2}}{4\pi}\sqrt{E}-\frac{1}{24}. \label{NoNe}
\end{eqnarray}

Our quantum results were obtained from the conformal map diagonalization
technique \cite{PR93}, and hereafter it is these values which are cited
as the reference set of eigenvalues. These reference eigenvalues are
calculated by using a very large Hamiltonian matrix (i.e. $dim=10000$) with
the result that the lowest 1000 eigenvalues have an accuracy no worse than
$10^{-10}$ of the mean level spacing. The numerical calculations for the
transfer matrix and its roots were performed on a Compaq Alpha 2100. Often it
was difficult to assign individual states to a semiclassical root at
$\lambda=0.15$ and $\lambda=0.1$ where the level repulsion and the $y$
reflection symmetry were often not sufficiently broken. Moreover since our
implementation of Bogomolny's method is never more precise than $0.5$ percent
of the mean level spacing it was an arbitrary choice in assigning to a
pair of semiclassical roots an even and odd eigenvalue pair in the quantum
spectrum separated by less than $0.1\%$ of the mean level spacing. Typically
if two semiclassical eigenvalues are very close to two quantum eigenvalues,
we assign one of the semiclassical eigenvalues to the quantum eigenvalue
closest to it and the other semiclassical eigenvalue to the remaining quantum
eigenvalue.

\begin{figure}
\epsfxsize=14cm
\epsfbox{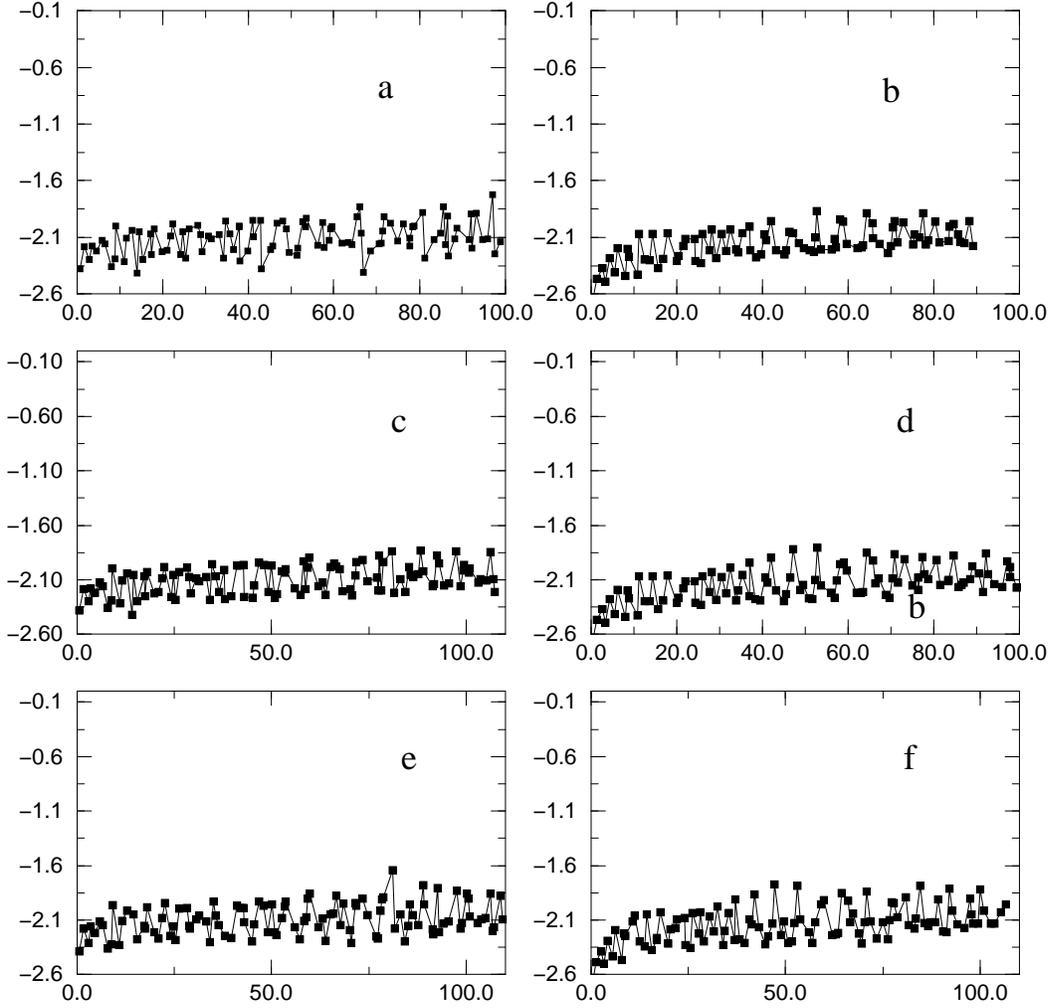}
\vspace{1cm}
%\narrowtext
\caption{Results of Bogomolny's transfer operator method for the hard
chaos cases for both even (a,c,e) and odd (b,d,f) eigenstates. We draw
$\alpha(E) = \log_{10}(|N(E_{sc})-N(E_{ex})|)$ versus $N(E_{ex})$. The top
figures (a and b) are for $\lambda=0.2$, the middle figures (c and d)
are for $\lambda=0.23$ and the bottom (e and f) are for
$\lambda=0.245$. All results are with dim(T)=350. The lines are drawn
for guiding the eye.} \label{fig1}
\end{figure}

\begin{figure}
\epsfxsize=14cm
\epsfbox{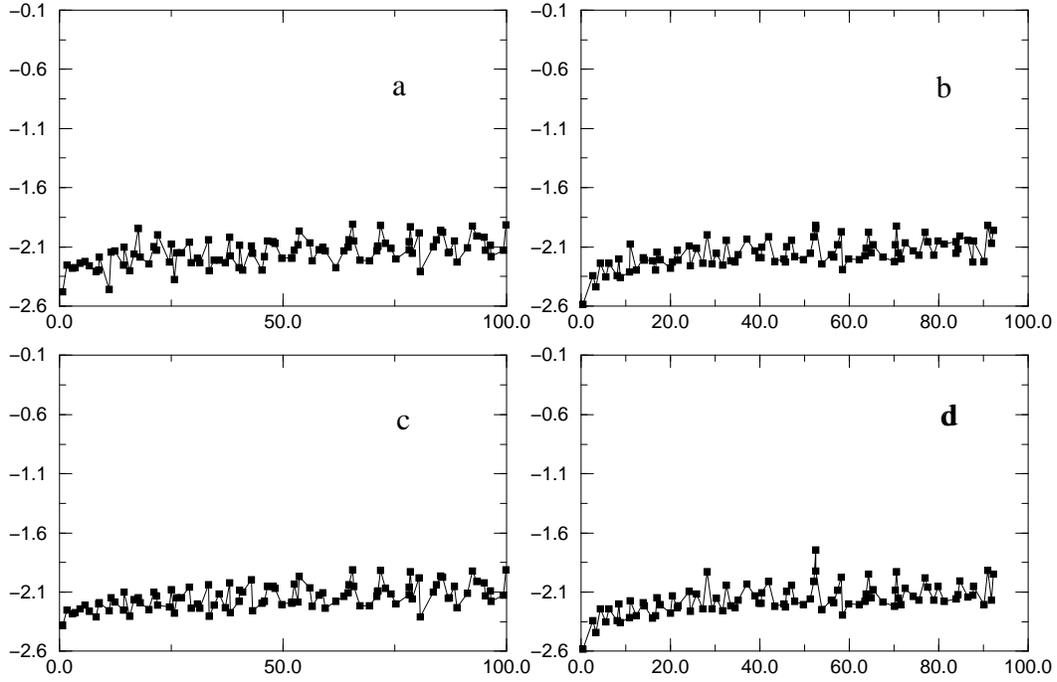}
\vspace{1cm}
%\narrowtext
\caption{The same as Fig. (1) but for
for $\lambda=0.1$ (a and b) and
$\lambda=0.15$ (c and d). Again a and c are for even states, b and d for
odd states.
All results are for dim(T)=350.  The lines are drawn for guiding the eye.}
\label{fig2}
\end{figure}

We show results for energies in the interval $E=(5,800)$  which includes
approximately the first two hundred states ($\approx$ one hundred odd and one
hundred even states) for the $\lambda$'s listed in the introduction. We also
report results for an energy interval in the deep semiclassical limit for
$\lambda=0.15$. We report all data in units of the mean level spacing
such that if $E_{sc}$ is the semiclassically predicted eigenvalue and
$E_{ex}$ is the quantum result, then the error we consider is

\begin{equation}
\alpha(E)=\log_{10}(|\Delta E|),\qquad \Delta E=N(E_{sc})-N(E_{ex})
\label{Error}
\end{equation}
In both Fig.~(\ref{fig1}) and Fig.~(\ref{fig2}) we plot
$\alpha(E)$ as a function of $N(E_{ex})$.

In order to maintain consistently good precision the parameter $b$ must be
kept constant over an entire energy range. On the other hand we consider a
constant matrix dimension and span an appreciable energy range. So in the 
data of Fig~(\ref{fig1}) and Fig.~(\ref{fig2}) we use a matrix of size $350$ 
for all $\lambda$ in the energy interval $(5,800)$ and in this range $b$ 
decreases from $150$ to $10$, a considerable change. The error fluctuates 
between $-1.8$ and $-2.5$, with the mean $\approx -2.1$ (which means that 
the error is about one percent of the mean level spacing) but does not 
change appreciably even with the large change in $b$. Again we emphasize 
that the reference eigenvalues come from the matrix diagonalization 
technique. 
A rough mean obtained by using the BIM without any semiclassical approximation 
has been shown to be of order $10^{-4}$ or $0.01\%$ of the mean level spacing
\cite{LRH98}. The semiclassical result obtained here is a factor of $100$ times 
less precise than the BIM if one uses $B\ge 10$. If one decreases $b$ 
to below ten, a deviation in accuracy does result and if one uses $b$ 
less than $3$, about $25\%$ of states in the spectrum are missed. 

We present our results for $\lambda\ge 0.2$ in Fig~(\ref{fig1}). The results
are equally good for both the softer chaos cases, $\lambda<0.2$ and shown in
Fig.~(\ref{fig2}). From these figures and additional results performed at
$\lambda=0.125,0.175,0.19$, it seems clear that there is no reason
to believe that the Bogomolny method is better in one classical regime 
than in
another. In Fig.~(\ref{fig1}) and Fig.~(\ref{fig2}) we see no deviation from
a characteristic mean for any particular state, indicating that the method is
unaffected by the classification of the wavefunction as being regular,
chaotic or mixed. This is indeed unlike the result expected from the GTF;
this semiclassical quantization tool would be invariably tedious to 
implement 
in the mixed regime. In the hard chaos regime, $\lambda=0.5$, and using a 
few short periodic orbits, the method has been shown to yield an 
accuracy of $8\%-10\%$ of the mean level spacing \cite{Whelan}.

It must be pointed out that the slight increase of the
semiclassical error with increasing sequence number in Fig. (1) and (2)
is not due to the semiclassical method, instead it results from
the decrease of $b$. Because in our numerical calculation, the dimension of
matrix $dim(T)$ is fixed at 350 as aforementioned, thus $b \sim 1/\sqrt{N}$, 
where $N$ is the sequence number.

\begin{figure}
\epsfxsize=14cm
\epsfbox{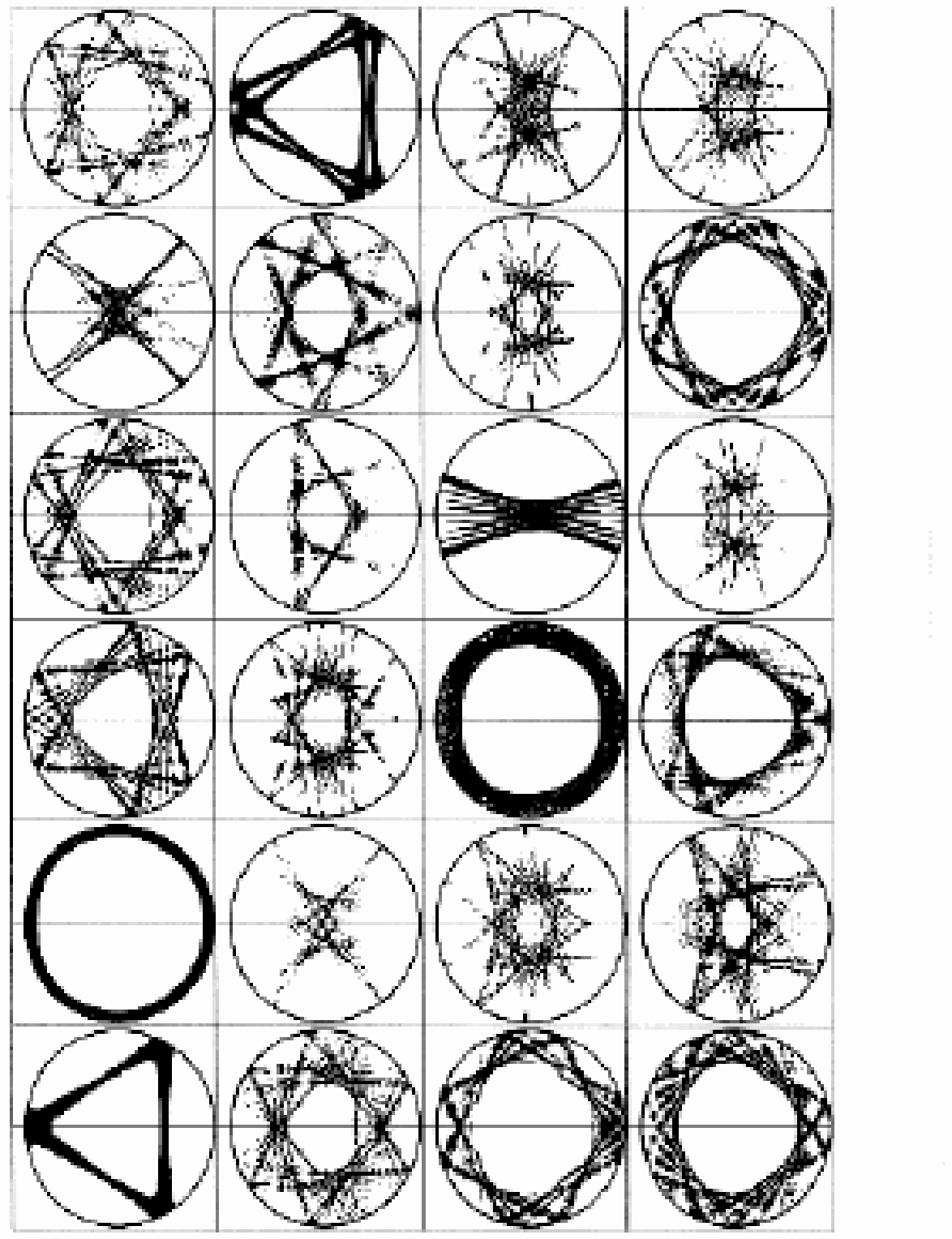}
\vspace{1cm}
%\narrowtext
\caption{The gallery of wavefunctions in coordinate space for 24 consecutive
eigenstates
starting from the $10,603$rd state. The order is left-right and top-down.}
\label{fig3}
\end{figure}

\begin{figure}
\epsfxsize=14cm
\epsfbox{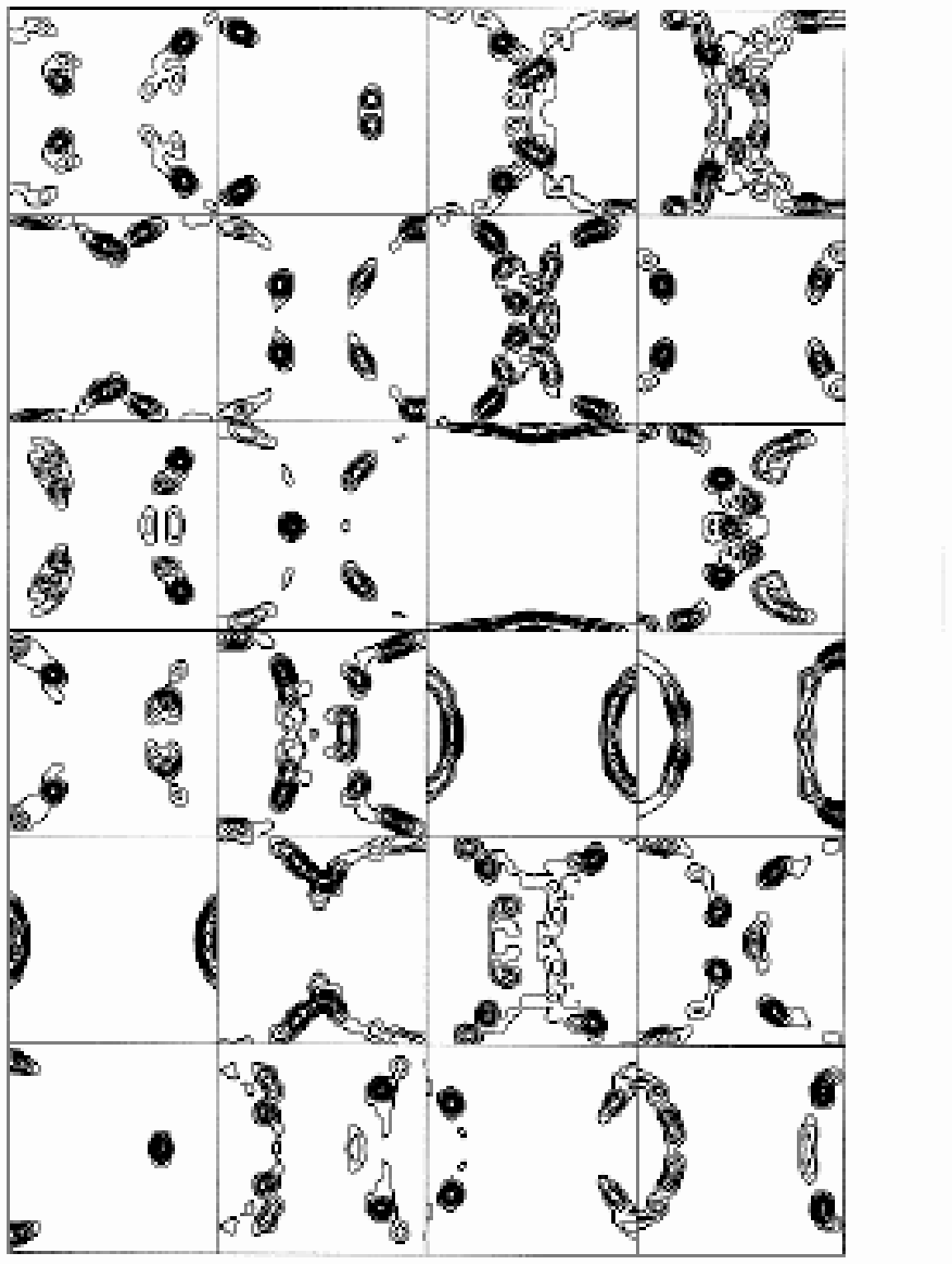}
\vspace{1cm}
%\narrowtext
\caption{Smoothed object $\rho_{SOS}$ of the wavefunctions in
Fig.~(\ref{fig3}). Again the order is left-right and top-down. The
abscissa is from $-1+\lambda$ to $1+\lambda$, and the vertical axis from
$-\sqrt{E}$ to $\sqrt{E}$. So the minimum quantum cell is about the size
$4/ \sqrt{E}$.} \label{fig4}
\end{figure}

To verify the method for soft chaos at high energy, we explore a small
energy range starting with the $10,603$rd even state with $dim(T)=1000$, and
thus $b\approx 3.2$. The corresponding energy range, $E=(81435,81600)$
includes $24$ odd and 24 even states. At these energies a classification of
each eigenstate in terms of a regular, chaotic or mixed description has been
obtained but again our calculations confirm that Bogomolny's scheme is
independent of the eigenstate's classification. We illustrate the $24$ even
wavefunctions in coordinate space in Fig.~(\ref{fig3}) and as a smoothed
Wigner function, see in Fig.~(\ref{fig4}). Their corresponding exact
eigenenergy, the eigenenergy obtained by Bogomolmny's scheme and their
classification as being chaotic, regular, or mixed are presented in Table I.
Compared with the average error at lower energy, one
finds that the error is also slightly increased. Again, this is due to
the decrease of $b$. This conclusion is different from that one obtained
by Prosen and Robnik\cite{PR93} for circular billiard with the torus
quantization. There they found that the semiclassical quantization error
increases with increasing energy.

The classification of eigenstates is based on the comparison of smoothed
projective Wigner function and that of the classical as used in our
previous work\cite{Li95A,Li95B}.
The Wigner function (of an eigenstate $\psi(u,v)$) defined in the full phase
space $(u,v,p_x,p_y)$ is
\begin{equation}
W({\bf q},{\bf p})=\frac{1}{(2\pi)^2}\int\;d{\bf X}^2\;
\exp{(-i{\bf p}{\bf X})}\psi({\bf q}-\frac{\bf X}{2})\;
\psi^{\dagger}({\bf q}+\frac{\bf X}{2})
\label{wigner}
\end{equation}
Here $\psi$ is a function of two variables ${\bf q}=(x,y)$ and
${\bf p}=(p_x,p_y)$. We have also put $\hbar=1$. In order to compare the
quantum Wigner functions with the classical Poincar\'e map we first choose a
SOS (not the boundary) and define a projection of $W({\bf q},{\bf p})$ onto
the SOS. The objective is to cast the Wigner function into a $2$-dim space of
one coordinate and its conjugate momentum. We take the SOS to be the line
$y=0$ and project the Wigner function of even states onto the $x-$axis,
\begin{equation}
\rho_{SOS}(x,p_x,y=0)=\int\;dp_y\;W(x,y=0,p_x,p_y),
\label{projection}
\end{equation}
which nicely reduces the number of integrations by one and is equal to
\begin{equation}
\rho_{SOS}(x,p_x)=\frac{1}{2\pi}\;\int\;da\;\exp{(iap_x)}\psi(x+\frac{a}{2},0)
\psi^{\dagger}(x-\frac{a}{2},0).
\label{smooth}
\end{equation}
As is well known that the Wigner function and its projections are not
positive definite and indeed one typically finds small and inconvenient but
nevertheless physical oscillations around zero which seriously obscure the
main structural features. Therefore in order to compare the classical and
quantal phase space structure we have smoothed the projection
Eq.~(\ref{smooth}) by using a normalized Gaussian kernel with a suitably
adapted dispersion, \cite{Takahashi,Leboeuf}.

In Fig.~(\ref{fig3}) we show the wavefunctions (in coordinate space)
corresponding to $24$ eigenstates of even parity beginning
with the $10,603$rd state. The reference eigenenergy was obtained by
diagonalization of a $32000\times 32000$ matrix. In Fig.~(\ref{fig4}) we 
plot the smoothed object $\rho_{SOS}$ in Eq.~(\ref{smooth}). The lowest 
contour shown is at the level of $0.15$ of the maximal value and the step 
size upwards is $0.15$ of the maximum.

\begin{figure}
\epsfxsize=14cm
\epsfbox{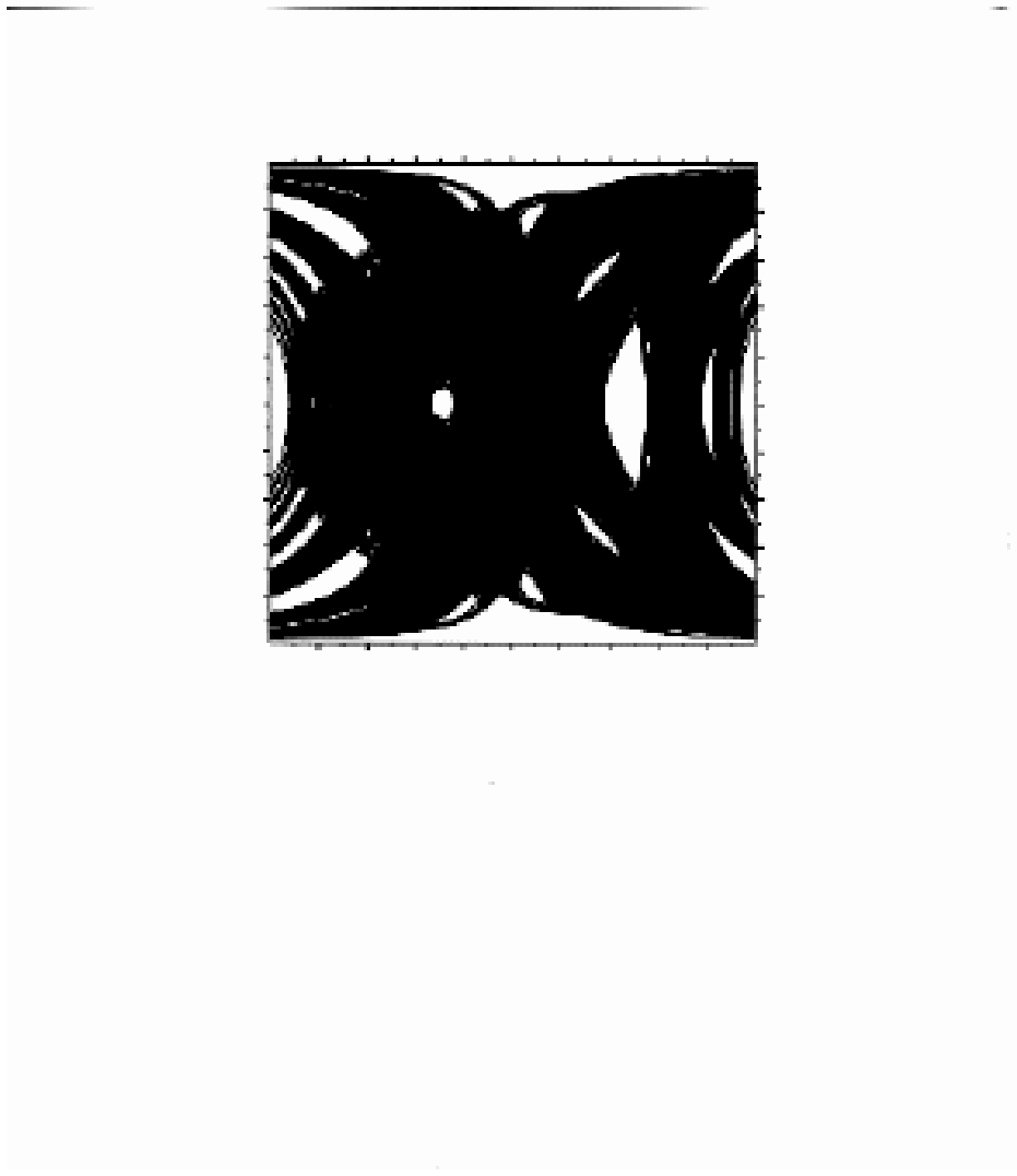}
\vspace{1cm}
%\narrowtext
\caption{Portraits of the classical phase space on the $y=0$ for
$\lambda=0.15$.}
\label{fig5}
\end{figure}

From a comparison of Fig.~(\ref{fig4}) and the classical phase space
on the $y=0$ SOS in Fig.~(\ref{fig5}) we can classify the wavefunction as
being either regular, chaotic or mixed. In Table I we note the
classification and the precision of Bogomolny's scheme. Again we hope that
this small energy range, but nevertheless representative, can help to
justify our conclusion that Bogomolny's scheme is independent of the
morphology of the eigenstate.

To examine the effect of increasing $b$ in the deep semiclassical limit,
we consider the states 10610 (mixed), 10611 (chaotic) and 10612 (regular).
The matrix dimension $dim(T^{osc})$ is increased from 700,
830 to 1000 which corresponds to the boundary node density $b$ is
approximately changed from 2.45, 2.91 to 3.44, respectively. The
results are shown in Table II. The enhancement of the accuracy is clearly  
shown in each of these three eigenstates of different classes.

\begin{figure}
\epsfxsize=14cm
\epsfbox{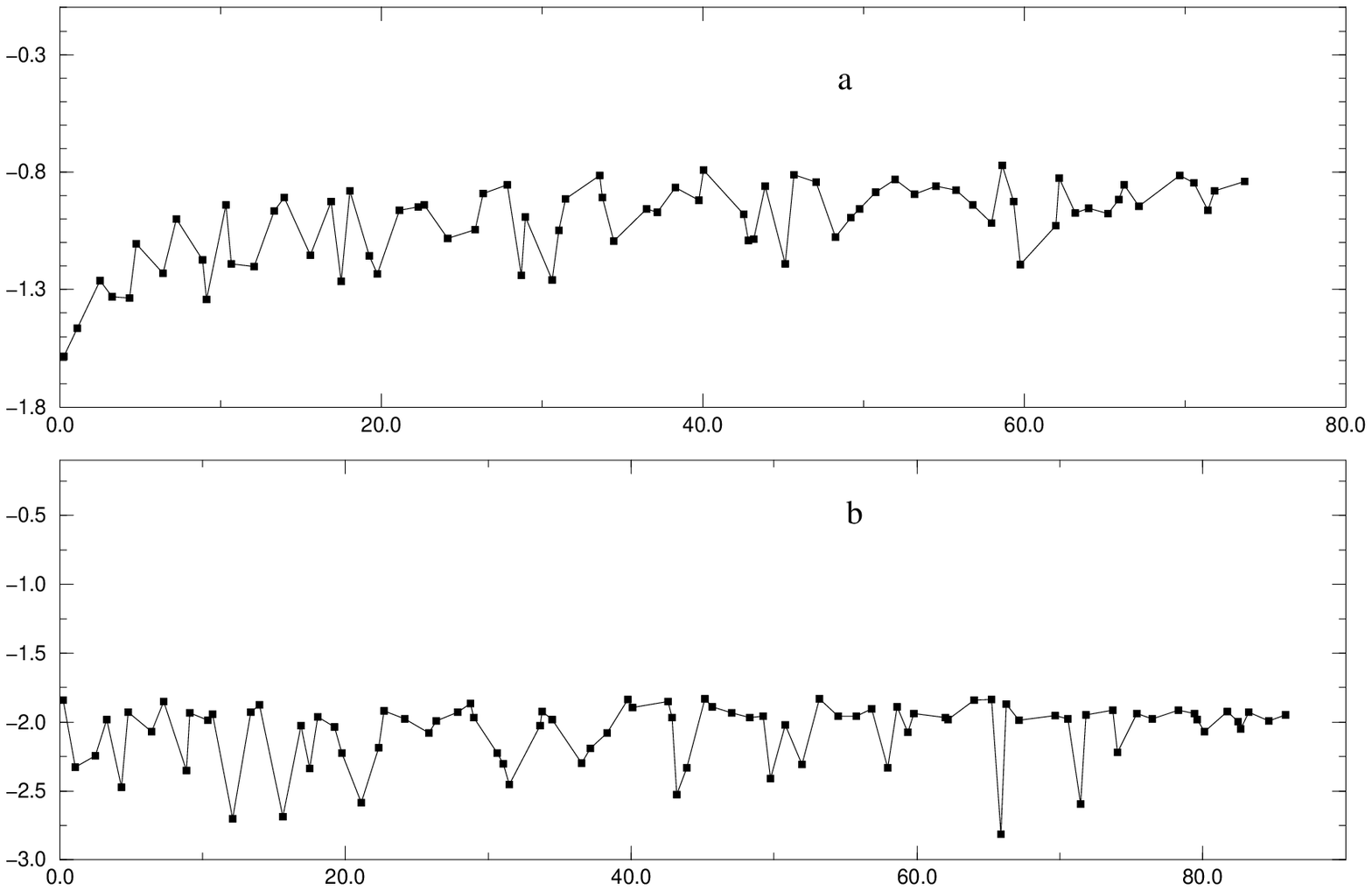}
\vspace{1cm}
%\narrowtext
\caption{Results of Bogomolny's transfer operator method for the odd-odd
parity eigenstates in the stadium billiard. The top figure (a) refers
to data using the quarter stadium and the lower figure (b) to the data
using the whole stadium and extracting only those energies close to
the reference eigenvalues.  The lines are drawn for guiding the eye.}
\label{fig6}
\end{figure}

As a further example, finally we consider the first hundred eigenstates of
the Bunimovich stadium billiard with odd-odd parity. The dimensions are the
following: the semicircle ends have radius $R=1$ and the half length of the
straight segment $a$ is fixed at one. We consider both the quarter stadium 
with Dirichlet boundary conditions on all four walls and the entire boundary 
which will give all four symmetry classes. The classical dynamics of this 
system has been shown to be ergodic \cite{bunimovich}.
The quantum reference eigenvalues were computed by using
the PWDM and are accurate to $10^{-4}$ of the mean level
spacing\cite{LRH98}. A comparison of the semiclassical result and the 
reference eigenvalues, seen to be comparable with those obtained for the 
Lima\c con billiard, are shown in Fig.~(\ref{fig6}).

\section{Conclusion and discussions}

We have studied the semiclassical quantization scheme of Bogomolny for two
quantum billiard systems, one of which is classically ergodic for all
physical parameters, the stadium billiard, and the other makes a smooth
transition from integrability to hard chaos as a parameter $\lambda$ is
changed.

We have studied the latter billiard for several values of $\lambda$
corresponding to both soft and hard chaos. We do not observe a dependence of
the numerical accuracy of the method on the classification of the wavefunction
as being either regular, chaotic or mixed. Furthermore with respect to a
reference set of eigenvalues, the accuracy of the semiclassical method is on
average about $1$ percent of the mean level
spacing. Here we use the eigenvalues obtained by the matrix diagonalization
technique as the reference set.

In the Bunimovich stadium billiard we obtain results for the first 100 states
of odd-odd parity. We find that the method obtains results that are very
much comparable with those for the Lima\c con billiard.

As some further applications of this work we may consider the calculation
of semiclassical eigenvalues in systems where ghost trajectories should be
included in the calculation of the transfer matrix. Examples would include any
billiard with a non-convex boundary (trajectories passing outside of the
billiard) or billiards such as the Sinai and annulus billiard.

We could also consider an application of Bogomolny's method in polygonal
billiard systems with one angle a rational multiple of $\pi$, a so-called
pseudo-integrable billiard \cite{BerryRichens,Bogomolny98a} where the
wavefunction shows multifractality\cite{Bogomolny98b}.

\section{Acknowlegements}
We would like to thank Dr. Niall D. Whelan and Dr. Pei-Qing Tong for
interesting
discussions, and Dr. Toma$\check{z}$ Prosen for providing some of the
reference eigenvalues of the Lima\c con billiard. We are also very
grateful to the anonymous referees for very useful comments and
suggestions. This work was supported in
part by the grants from the Hong Kong Research Grants Council (RGC) and the
Hong Kong Baptist University Faculty Research Grants (FRG).

\newpage

{\bf Table I.} The classification of the 24 consecutive high-lying even
states of the Lima\c con billiard at $\lambda=0.15$. By comparing the
smoothed function $\rho_{SOS}$ with the classical phase portrait (fig 5),
we can distinguish the states as chaotic (C), regular (R) and
mixed (M). The semiclassical eigenenergy obtained by Bogolmony's scheme
is given and compared with the reference quantum eigenenergies. The error
is measured in the unit of the mean level spacing. For all these states
we used a $1000$ dimensional array ($b\approx 3.51$) except for the $10,614,
10,622$th and $10,623$th states. Here we had to use a larger matrix of
dimension $1200$, ($b\approx 4.22$), since by using smaller $b$, these
two levels are always missed.
%\vspace{5mm}\\
\begin{tabular}{|l|c|c|c|c|}\hline
& & & &\\
$N$ & Classification  & $E_{ex}$&
$E_{sc}$& $|\Delta E|$\\[2.0ex]\hline & & & & \\
10603& c & $81,361.69$&$81,361.88$&2.4e-2\\[1ex]
10604& c & $81,369.91$&$81,370.10$&2.4e-2\\[1ex]
10605& c & $81,372.41$&$81,372.51$&1.3e-2\\[1ex]
10606& c & $81,377.38$&$81,377.52$&1.8e-2\\[1ex]
10607& c & $81,382.99$&$81,382.77$&2.8e-2\\[1ex]
10608& c & $81,398.52$&$81,398.39$&1.6e-2\\[1ex]
10609& c & $81,408.35$&$81,408.50$&1.9e-2\\[1ex]
10610& m & $81,435.34$&$81,435.57$&2.9e-2\\[1ex]
10611& c & $81,441.86$&$81,442.05$&2.4e-2\\[1ex]
10612& r & $81,451.87$&$81,452.47$&7.5e-2\\[1ex]
10613& r & $81,455.08$&$81,455.35$&3.4e-2\\[1ex]
10614& c & $81,458.73$&$81,459.03$&3.8e-2\\[1ex]
10615& r & $81,462.10$&$81,462.28$&2.3e-2\\[1ex]
10616& m & $81,472.86$&$81,472.97$&1.4e-2\\[1ex]
10617& c & $81,501.51$&$81,501.76$&3.1e-2\\[1ex]
10618& c & $81,504.34$&$81,504.65$&3.9e-2\\[1ex]
10619& r & $81,507.96$&$81,508.23$&3.9e-2\\[1ex]
10620& c & $81,511.94$&$81,512.05$&1.4e-2\\[1ex]
10621& c & $81,512.74$&$81,512.95$&2.6e-2\\[1ex]
10622& c & $81,534.52$&$81,535.00$&6.0e-2\\[1ex]
10623& r & $81,536.83$&$81,537.03$&2.5e-2\\[1ex]
10624& c & $81,538.91$&$81,539.09$&2.3e-2\\[1ex]
10625& r & $81,543.85$&$81,544.07$&2.8e-2\\[1ex]
10626& c & $81,558.03$&$81,888.24$&2.6e-2\\[1ex]
\hline
\end{tabular}

\newpage
{\bf Table II.} The dependence of the eigenenergies with
the boundary nodes density $b$ for three different classes of
eigenstates, regular, mixed and chaotic.
The matrix dimension $dim(T^{osc})$ is increased from 700,
830 to 999. The boundary nodes density $b$ is approximately
changed from 2.45, 2.91 to 3.44, respectively.  The enhancement of the
accuracy is demonstrated in each of these three eigenstates of
different classes.
\bigskip
\bigskip
\bigskip

\begin{tabular}{|l|c|c|c|c|c|}\hline
& & & & &\\
$N$ & Classification  & $E_{ex}$ &
$E_{sc} (b=2.45)$& $E_{sc} (b=2.91)$& $E_{sc} (b=3.44)$\\[2ex]\hline
& & & & &\\
10610& m & $81,435.34$ & $81,435.68$ & $81,435.62$ & $81,435.58$\\[1ex]
10611& c & $81,441.86$ & $81,442.15$ & $81,442.10$ & $81,442.07$ \\[1ex]
10612& r & $81,451.87$ & $81,452.52$ & $81,452.50$ &  $81,452.47$ \\[1ex]
\hline
\end{tabular}

%\end{multicols}

\begin{thebibliography}{9}
\bibitem{GutzB}
M.\ C.\ Gutzwiller, {\it Chaos in Classical and Quantum Mechanics},
(New York: Springer 1990).

\bibitem{BohigasLesH91}
{\it Chaos and Quantum Systems} (Proc.NATO ASI Les Houches Summer School),
ed.\ M-J Giannoni, A.\ Voros and J.\ Zinn-Justin (ed. Amsterdam: Elseivier),
pg.\ 87.

\bibitem{Casati93}
G.\ Casati and B.\ V.\ Chirikov, {\it Quantum Chaos},
(Cambridge University Press, Cambdridge 1995).

\bibitem{Heller84}
E.\ J.\ Heller,
Phys.\ Rev.\ Lett.\ {\bf 53}, 1515  (1984).

\bibitem{Bogomolny88}
E.\ B.\ Bogomolny,
Physica {\bf D31 }, 169 (1988).

\bibitem{Berry89}
M.\ V.\ Berry,
Proc. R. Soc. A {\bf 423}, 219 (1989).

\bibitem{Fishman93}
O.\ Agam and S.\ Fishman, J. Phys. A {\bf 26}, 2113 (1994); Phys.
Rev. Lett. {\bf 73}, 806(1994)

\bibitem{Li97}
B.\ Li,
Phys. Rev. E {\bf 55}, 5376 (1997).

\bibitem{Li98}
B.\ Li and B.\ Hu,
J.\ Phys.\ A\  {\bf 31}, 483 (1998).

\bibitem{Heller98}
L.\  Kaplan and E.\ J.\ Heller,
Ann.\ Phys.\ (NY) {\bf 264}, 171 (1998).

\bibitem{HPWDM}
E. J. Heller, in Ref.\cite{BohigasLesH91}, p. 548.

\bibitem{Li94}
B.\ Li and M.\ Robnik,
J.\ Phys.\ A  {\bf 27}, 5509 (1994).

\bibitem{Berry84}
M.\ V.\ Berry and M.\ Wilkinson,
Proc.\ R.\ Soc.\ London\ A {\bf 392}, 15 (1984).

\bibitem{Boasman}
P.\ A.\ Boasman,
Nonlinearity {\bf 7}, 5509 (1994).

\bibitem{LRH98}
B.\ Li,\ M.\ Robnik, and B.\ Hu,
Phys.\ Rev.\ E.\, {\bf 57}, 4095 (1998).

\bibitem{Robnik83B}  M.\ Robnik,
J.\ Phys.\ A {\bf 17}, 1049 (1984).

\bibitem{Berry86}
M.\ V.\ Berry and M.\ Robnik,
J.\ Phys.\ A\ {\bf 19}, 649 (1986).

\bibitem{PR93} T.\ Prosen and M.\ Robnik,
J.\ Phys.\ A\ {\bf 26}, 2371 (1993);
{\it ibid.} {\bf 27}, 8059 (1994).

\bibitem{Bohigas93}
O.\ Bohigas, D.Boos\'e, R. Egydio de Carvalho, and V. Marvulle,
Nucl.\ Phys.\ A \ {\bf 560}, 197 (1993).

\bibitem{uzy1}
E.\ Doron \ and U.\  Smilansky,
Nonlinearity, {\bf 5}, 1055 (1992);
H.\ Schanz and U. \ Smilansky,
Chaos, Fractals and Solitons {\bf 5}, 1289 (1995).

\bibitem{Prosen96}
T.\ Prosen,
J.\ Phys.\ A, {\bf 27}, L709 (1994); {\bf 28}, L349, 4133 (1995); Physica
{\bf D91}, 244 (1996).

\bibitem{VS95}
E.\ Vergini and M.\ Saraceno, Phys. Rev. E {\bf 52}, 2204 (1995).

\bibitem{truncate}
M.\ V.\ Berry and J.\ P.\ Keating, Proc. R. Soc. London {\bf A437}, 151
(1992).

\bibitem{cycleexp}
P.\ Cvitanovi\'c, Phys. Rev. Lett., {\bf 61},  2729 (1988);
R. Artuso, E. Aurell, and P. Cvitanovi\'c, Nonlinearity {\bf 3}, 325, 361
(1990).

\bibitem{Aurich}
R.\ Aurich, C.\ Matthies, M.\ Sieber, and F.\ Steiner,
Phys.\ Rev.\ Lett.\ {\bf 68}, 1629 (1992).

\bibitem{GTF2}
A M O de Almeida and J H Hannay,
J. Phys. A, {\bf 20}, 5873 (1987);
P. Gaspard and D. Alonso, Phys. Rev. A {\bf 47}, R3468 (1993); G. Vattay
and P. E.  Rosenqvist, Phys. Rev. Lett. {\bf 76}, 335
(1996); G. Vattay, {\it ibid.} {\bf 76}, 1059 (1996);
H. Schomerus and M. Sieber,
J. Phys. A, {\bf 30}, 4537 (1997).

\bibitem{Bogomolny92} E.\ B.\ Bogomolny,
Nonlinearity {\bf 5}, 805 (1992).

\bibitem{semito}
P.\ Cvitanovi\'c and G.\ Vattay,
Phys.\ Rev.\ Lett.\ {\bf 71}, 4138 (1993); P.\ Cvitanovi\'c {\it et al.},
Chaos {\bf 3} (4), 619 (1993); B.\ Georgeot and R.\ E.\ Prange,
Phys.\ Rev.\ Lett.\ {\bf 74}, 2851, 4110 (1995).

\bibitem{Lauritzen}
B.\ Lauritzen,
Chaos {\bf 2}, 409 (1992).

\bibitem{Tong}
P.\ Tong and D.\ Goodings,
J.\ Phys.\ A \ {\bf 29}, 4065 (1997).

\bibitem{Goodings97}
N.\ C.\ Snaith and D.\ A.\ Goodings,
Phys.\ Rev.\ E {\bf 55}, 5212 (1997).

\bibitem{Goodings98}
D.\ A.\ Goodings and N.\ D.\ Whelan
J.\ Phys.\ A \ {\bf 31}, 7521(1998).

\bibitem{Bogomolny92B}
E.\ B.\ Bogomolny and M.\ Caroli,
Physica {\bf D67}, 88 (1993).

\bibitem{GoodingsN}
T.\ Szeredi,\ J.\ Levebvre and D.\  A. \ Goodings,
Nonlinearity {\bf 7}, 1463 (1994);
Phys.\ Rev.\ Lett. {\bf 71}, 2891 (1993).

\bibitem{Haggerty}
M.\ R.\ Haggerty,
Phys.\ Rev.\ E {\bf 52}, 389 (1995).

\bibitem{Robnik83A}
M.\ Robnik, J.\ Phys.\ A \ {\bf 16}, 3971 (1983).

\bibitem{Hayli87}
A.\ Hayli,\ T.\ Dumont,\ J.\ Moulin-ollagier, and J.\ M.\ Strelcyn,
J.\ Phys.\ A \ {\bf 20}, 3237 (1987).

\bibitem{Markarian}
R.\ Markarian,
Nonlinearity {\bf 6}, 819 (1993).

\bibitem{percival}
I.\ C.\ Percival,
J.\ Phys.\ B {\bf 6}, L229 (1973).

\bibitem{BerryRobnik}
M.\ V.\ Berry and M.\ Robnik,
J.\ Phys.\ A\ {\bf 17}, 2413 (1984).

\bibitem{Li95A}
B.\ Li and M.\ Robnik,
J.\ Phys.\ A \ {\bf 28}, 2799 (1995);
B. Li and M. Robnik, chaos-dyn/9501022.

\bibitem{Li95B}
B.\ Li and M.\ Robnik,
J.\ Phys.\ A\ {\bf 28}, 4843 (1995).

\bibitem{bunimovich}
L.\ A.\ Bunimovich,
Funct.\ Anal.\ Appl. {\bf 8}, 254(1974);
Commun.\ Math.\ Phys. {\bf 65}, 295(1979).

\bibitem{Whelan}
H.\ Bruus and N.\ D.\ Whelan,
Nonlinearity {\bf 9}, 1023 (1996) .

\bibitem{Takahashi}
K.\ Takahashi,
Prog.\ Theor.\ Phys.\ Suppl. {\bf 98}, 109 (1989).

\bibitem{Leboeuf}
P.\ Leboeuf and M.\ Saraceno,
J.\ Phys.\ A \ {\bf 23}, 1745 (1990).

\bibitem{Berry77}
M.\ V.\ Berry,
J.\ Phys.\ A \ {\bf 10}, 2083 (1977).

\bibitem{BerryRichens}
P.\ J.\ Richens and M.\ V.\ Berry,
Physica {\bf D 2}, 495 (1981).

\bibitem{Bogomolny98a}
E.\ B.\ Bogomolny, U.\ Gerland and C.\ Schmit, Phys. Rev. E {\bf 59}, R1315
(1999).

\bibitem{Bogomolny98b}
E.\ B. Bogomolny, Private communication.

\end{thebibliography}
\end{document}